\documentclass{aa} 
\usepackage{graphicx}
\usepackage{amsmath}
\usepackage[utf8]{inputenc}
\usepackage{wrapfig}
\usepackage{fancyhdr}
\usepackage{color}
\usepackage{longtable}
\usepackage{natbib}
\usepackage{changepage}
\usepackage{textcomp}
\usepackage{appendix}
\usepackage{txfonts}
\usepackage{multicol}

\def\lesssim{\mathrel{\hbox{\rlap{\hbox{\lower3pt\hbox{$\sim$}}}\hbox{\raise2pt\hbox{$<$}}}}}

\def\utw{\smash{\rlap{\lower5pt\hbox{$\sim$}}}}
\def\udtw{\smash{\rlap{\lower6pt\hbox{$\approx$}}}}

\begin{document}

\title{Small Bodies: Near and Far\ Database for thermal infrared observations of small bodies in the Solar System}


\author{R\'obert Szak\'ats
          \inst{1}
          \and
        Thomas M\"uller \inst{2} \and
        V\'{i}ctor Al\'i-Lagoa \inst{2} \and
        G\'abor Marton \inst{1} \and
        Anik\'o Farkas-Tak\'acs \inst{1} \and
        Evelin B\'anyai \inst{1} \and
        Csaba Kiss \inst{1}
          }

\institute{Konkoly Observatory, Research Centre for Astronomy and Earth Sciences, H-1121 Budapest, Konkoly Thege Miklos ut 15-17, Hungary\\
              \email{szakats.robert@csfk.mta.hu}
         \and
             Max-Planck-Institut f\"ur extraterrestrische Physik, Giesenbachstrasse, Garching, Germany\\
             }

\date{Received: 2019. June 21.; accepted 2019. December 31.}

 
\abstract{In this paper, we present the Small Bodies: Near and Far (SBNAF) Infrared Database, an easy-to-use tool intended to facilitate the modelling of thermal emission of small bodies of the Solar System. Our database collects measurements of thermal emissions for small Solar System targets that are otherwise available in scattered sources and provides a complete description of the data, including all information necessary to perform direct scientific analyses and without the need to access additional external resources. This public database contains representative data of asteroid observations of large surveys (e.g. AKARI, IRAS, and WISE) as well as a collection of small body observations of infrared space telescopes (e.g. the Herschel Space Observatory) and provides a web interface to access this data (https://ird.konkoly.hu).We also provide an example for the direct application of the database and show how it can be used to estimate the thermal inertia of specific populations, e.g. asteroids within a given size range. We show how different scalings of thermal inertia with heliocentric distance (i.e. temperature) may affect our interpretation of the data and discuss why the widely-used radiative conductivity exponent ($\alpha$\,=\,--3/4) might not be adequate in general, as suggested in previous studies.}
\authorrunning{Szak\'ats et al.}
\titlerunning{"The Small Bodies: Near and Far" Infrared Database} 
\maketitle
%

\section{Introduction}
\label{intro}

The field of analysing and modelling the thermal emission of asteroids has experienced substantial growth in the last decade, mainly thanks to the improved availability of thermal emission data \citep{Delbo2015} and it will continue thanks to the rising availability of shape models \citep{Durech2015} and spatially resolved in situ thermal emission data from space instruments \citep[the Hayabusa-2 and OSIRIS-REx missions,][]{Watanabe,Enos}.

The typical subsolar temperatures of near-Earth asteroids are $\sim$300\,K, and $\sim$200\,K in the main belt. The thermal emission of these asteroids peak in the mid-infrared (10-20$\mu$m). Although these wavelengths are partly available from the ground, the vast majority of these observations are performed by space instruments \citep[e.g. the Akari Space Telescope and the WISE/NEOWISE surveys][]{Usui2011,Masiero2011}.
Despite their warm surface temperatures, the modelling of near-Earth asteroids benefited from far-infrared observations too, such as, the thermal modelling of (101955) Bennu, (308635) 2005\,YU$_{55}$, (99942) Apophis, and (162173) Ryugu \citep{Muller2012,Muller2013,Muller2014c,Muller2017}. 
Beyond Jupiter and in the transneptunian region, the typical surface temperatures drop from $\sim$100\,K to 30--50\,K and the corresponding emission with a peak in the far-infrared (50-100\,$\mu$m) is almost exclusively observed from space. Centaurs and TNOs were suitably observable with the far-infrared detectors of the Spitzer Space Telescope \citep[mostly at  70\,$\mu$m,][]{Stansberry2008} and Herschel Space Observatory \citep[70, 100 and 160\,$\mu$m,][]{Muller2009,Muller2018,Muller2019}.
Recent reviews on the thermal emission of asteroids and more distant objects by space instruments can be found in \cite{Muller2019}, \cite{Mainzer2015}, and references therein.

While the original primary goal of asteroid thermal infrared measurements was to derive diameters and albedos, the improvement of detector sensitivity and the availability of multi-epoch observations not only offer more precise diameters and albedos, but also allow fort the derivation of other physical properties, such as thermal inertia and surface roughness, and, consequently, such features as the porosity and physical nature of the regolith layer. Thermal emission measurements also provide information on the YORP effect \citep{Vokrouhlicky2015} and they can constrain the spin-axis orientation \citep[see e.g.][]{Muller2012,Muller2013,Muller2017}, as well as the temperature evolution that can alter the surface or subsurface composition as well as help in shape and spin modelling \citep{Durech2017}, while also providing information on diurnal temperature variations that may lead to thermal cracking and, thereby, produce fresh regolith \citep{Delbo2014}. 

The interpretation of the thermal emission measurements is rather complex, as the measured flux densities are strongly dependent on the epoch of the observations through the heliocentric distance of the target, the distance between the target and the observer, phase angle, aspect angle, rotational phase, and also on the thermal and surface roughness properties. The actual thermal characteristics are not fundamental properties, as, for example, the thermal inertia is a function of thermal conductivity, heat capacity, and density, while thermal conductivity and heat capacity both depend on the local temperature \citep[see][and references therein]{Keihm1984,Keihm2012,Delbo2015, Marsset2017, Rozitis2018}. Researchers working on these types of interpretations need to collect and process all of this auxiliary information to correctly interpret the thermal emission measurements individually for all type of instruments.

The primary goal of the Small Bodies: Near and Far (SBNAF) Infrared Database (hereafter, IRDB) is to help scientists working in the field of modelling the thermal emission of small bodies by providing them with an easy-to-use tool. Our database collects available thermal emission measurements for small Solar Systems targets that are otherwise available in scattered sources and gives a complete description of the data, including all the information necessary to perform direct scientific calculations and without the need to access additional external resources. 
The IRDB provides disk-integrated, calibrated flux densities based on careful considerations of instrument- and project-specific calibration and processing steps. These multi-epoch, multi-wavelength, multi-aspect data allow for a more complex thermophysical analysis for individual objects (e.g. using more sophisticated spin-shape solutions) or samples of objects. It will also allow for the combination of remote data with close-proximity data for the same target. In addition to answering direct scientific questions related  to, for example, thermal inertia and other surface properties of the targets, it will also help in establishing celestial calibrators for instruments working in the thermal infrared regime, from mid-IR to submm wavelengths \citep[see e.g.][]{Muller2014b}.

Early versions of the database were used in several studies. \citet{Marciniak2018,Marciniak2019} performed a modelling of long-period and low-amplitude asteroids and dervied the thermal properties of slowly rotating asteroids; \citet{Muller2017} obtained the spin-axis orientation of Ryugu using data from several infrared space and ground based instruments; and the 3D shape modelling and interpretation of the VLT/Sphere imaging of (6) Hebe was also supplemented by data from the infrared database \citep{Marsset2017}. Reconsiderations of the thermal emission of Haumea \citep{Muller2018b}, constrained with the occultation results \citep{Ortiz2017} was also performed using multi-epoch, multi-mission data from the SBNAF IRDB. 

In the present version of the database, we included flux densities only at those wavelengths which are expected to be purely thermal, with a negligible contribution from reflected solar radiation. This excludes, for example, the two short wavelength filters of WISE, W1 and W2 (3.4 and 4.6$\mu$m).

Our final aim is to include thermal data for all Solar System small bodies which have been detected at thermal infrared wavelengths. The initial version of the IRDB has been created in the framework  of the Horizon 2020 project known as Small Bodies: Near and Far \citep[COMPET-05-2015/687378,][]{Muller2018}. 

\textbf{ Researchers working on small body thermal emission topics are encouraged to submit their own processed and calibrated thermal infrared observations to the SBNAF IRDB. These will be  transformed to the database standards, supplemented with auxiliary data and made available to the planetary science community.}

The structure of the paper is the following. In Sect.~\ref{sect:obs}, we summarize the scheme of the SBNAF Infrared Database and give a detailed summary of its resources. In Sect.~\ref{sect:aux}, we describe the auxiliary data that supplement the core data. A detailed list of the database fields and SQL examples of the possible queries are provided in Sect.~\ref{sect:database}. In Sect.~\ref{sect:albedos}, we investigate the impact of albedos obtained from different resources on the colour correction values we calculate as auxiliary data and we provide an example application to demonstrate the capabilities of the SBNAF Infrared Database, and derive thermal inertias of some specific asteroid populations. Future aspects related to the database, including the submission of data to the IRDB, are discussed in Sect.~\ref{sect:updates}.  

\section{Thermal infrared observations of asteroids and transneptunian objects\label{sect:obs}}

\subsection{Database scheme \label{sect:scheme}}

\begin{figure*}[ht!]
  \centering
  \includegraphics[width=1.0\linewidth]{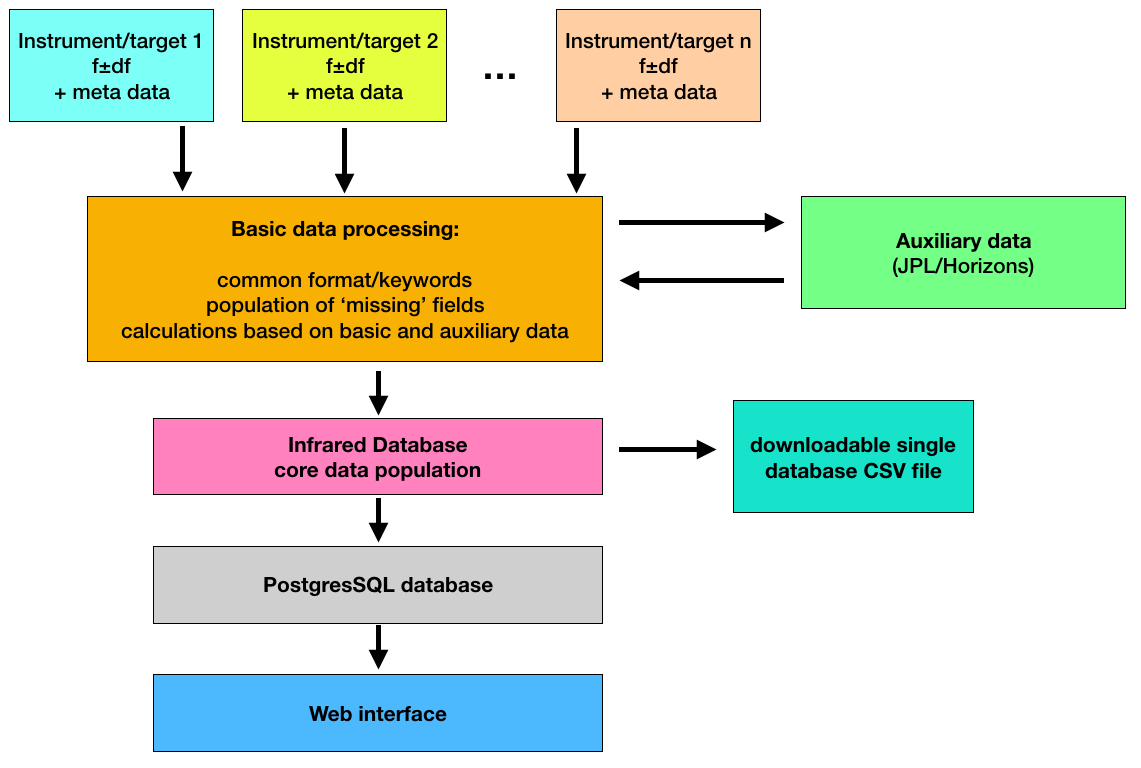}
  \caption{Outline of the processing of data from the basic entries (catalogues and target-specific papers) to the web-interface of the SBNAF infrared database. 
  }\label{fig:outline}
\end{figure*}

The main entries in our database are the (calibrated) infrared flux densities and the corresponding flux density error (denoted as $f\pm df$ in the outline figure Fig.\ref{fig:outline}), supplemented with observational meta data -- object identifier, observatory, measurement identifier, instrument-band-filter-observing mode, start and end time of the observations, duration, measured in-band flux (calibrated, aperture-/beam-corrected, non-linearity/saturation-corrected, etc.). Raw flux densities or errors and observational meta data are typically available in the catalogues or target-specific papers where we take our basic data from. These papers/catalogues are listed in Sect.~\ref{sect:keyref}. All these flux densities are processed (e.g. converted to [Jy] from magnitudes) and brought to a common format along with all meta data in our processing (see also Fig.\ref{fig:outline}). 

The final aim is to include data of near-Earth, main-belt, and transneptunian objects with significant thermal measurements from different satellite missions (IRAS, MSX, ISO, AKARI, Herschel, WISE/NEOWISE) and also from the ground. In the present, first public release, we include IRAS, MSX, Akari, Herschel, and WISE/NEOWISE observations. A more detailed description of these missions and references to available infrared data are given in Sect.~\ref{sect:keyref}.
The current version of the catalogue contains 169\,980 entries (see Table \ref{table:instruments}.).

\begin{table*}
\small
\centering
\begin{tabular}{cccccr}
\hline
Mission & instrument & filters & observing mode & N$_{obs}$\\
\hline\hline
AKARI & IRC-NIR     &  N4            & IRC02                & 1 \\
                                 &      IRC-MIR-S   & S7, S9W, S11   & survey, IRC02, IRC11 & 6955 \\
                         &      IRC-MIR-L   & L15, L18W, L24 & survey, IRC02, IRC51 & 13824  \\
\hline
HSO             &       PACS &  blue,green,red  & chop-nod, scan map        & 1852 \\
\hline
MSX                     &       & MSX\_A,MSX\_C,MSX\_D,MSX\_E & survey              & 901 \\
\hline\
IRAS            &   & IRAS12,IRAS25,IRAS60,IRAS100 & survey             & 25064  \\
\hline
WISE    &   & W3,W4 & survey                                    & 121383 \\
\hline
\end{tabular}
\caption{\small List of observatories or missions, instruments, filters, possible observing modes, and the number of measurements with a specific instrument, in the present version of the Infrared Database. Instruments on low-Earth orbits, such as AKARI, MSX and IRAS are referred to as geocentric (JPL code '500@399'), and we used the JPL code '500@-163' for WISE and '500@-486' for Herschel. \label{table:instruments}}
\end{table*}


To fully utilize these measurements, we collect auxiliary data for the observations from external sources. These data are partly stored as additional useful entries (e.g. orbital elements and coordinates from JPL/Horizons) or used to calculate quantities that are necessary for the correct interpretation of the measurements (e.g. colour correction). A list of quality comments or other relevant flags are also included. 
The IR database is created in such a way that all available thermal measurements of our selected targets are easily accessible through a simple web interface: {\bf\url{ https://ird.konkoly.hu}}.

\subsection{Key references and sources of data \label{sect:keyref}}

Below we give a summary of the references and sources of asteroid flux densities related to each mission and telescope. 
A simple reference code is also given for those papers or catalogues that presented flux densities used for our database. These codes are listed in the {\sl  'documents\_references'} field of the database for each measurement. 

\paragraph{IRAS:}
A general description of the Infrared Astronomical Satellite (IRAS) mission can be found in \cite{Neugebauer1984}. 
A detailed summary of the IRAS mission is given in the IRAS Explanatory Supplement, available at the NASA/IPAC Infrared Science Archive \textbf{\footnote{http://irsa.ipac.caltech.edu/IRASdocs/exp.sup/}}, that also covers calibration issues (point source calibration, estimated accuracy, bright source problems, colour correction). Asteroid fluxes are obtained from The Supplemental IRAS Minor Planet Survey from \cite{Tedesco2002} [{\bf T02IRAS}]. 

\paragraph{MSX:}

\cite{Mill1994} and \cite{Milletal1994} provide an overview of the spacecraft, its instruments, and scientific objectives, and \cite{Price1995,Price1998} a general description of the astronomy experiments. More details on the astronomy experiments and the influence the spacecraft design has on these experiments can be found in \cite{Price2001}. Asteroid fluxes are obtained from The Midcourse Space Experiment Infrared Minor Planet Survey catalogue from \cite{Tedesco2002x} [{\bf T02MSX}]. 

\paragraph{AKARI:}

The AKARI mission is described in \cite{Murakami}. 
Minor planet flux densities are obtained from the AKARI Asteroid Flux Catalog Ver.1
\footnote{\url{https://www.ir.isas.jaxa.jp/AKARI/Archive/Catalogues/Asteroid\_Flux\_V1/}}
(Release October 2016), referred to as [{\bf AKARIAFC}] in the IRDB. The catalogue contains data from the all-sky survey by \cite{Usui2011}, slow-scan observation by  \cite{Hasegawa2013}, and pointed observations of (25143)~Itokawa and (162173)~Ryugu by  \cite{Muller2014,Muller2017}. These 'non-survey' modes have their special flags in the 'obsmode' fields (IRC02: pointed observations; IRC11/IRC51: slow scan), as defined in the AKARI Asteroid Flux Catalogue. 

\paragraph{Herschel:}
The Herschel Space Observatory mission is summarized in \cite{Herschel}. The  Photometer Array Camera and Spectrometer (PACS) instrument is described in \cite{PACS}. The photometric calibration of PACS is discussed in \cite{Nielbock2013} (chop-nod photometric observing mode) and in \cite{Balog2014} (scan-map photometric mode).

Flux densities of Solar System small bodies are obtained from selected publications of near-Earth asteroids and Centaurs/transneptunian objects.

\smallskip
\begin{table*}[ht!]
    \centering
    \scriptsize
    \begin{tabular}{p{3.2cm}p{1.3cm}p{8cm}}
    \hline
     \it Near-Earth asteroids & & \\
     \hline
     M\"uller et al. (2012) & {\bf [M12]} & (101955) Bennu \\
     M\"uller et al. (2017) & {\bf [M13]} & (308625) 2005\,YU55 \\
     M\"uller et al. (2017) & {\bf [M14]} & (99942) Apophis \\
     M\"uller et al. (2017) & {\bf [M17]} & (162173) Ryugu \\
    \hline
    \multicolumn{3}{l}{\it Centaurs and transneptunian objects}  \\
    \hline
     Duffard et al. (2014) & {\bf [D14]} & 16 Centaurs \\
    Fornasier et al. (2013) & {\bf [F13]} & (2060) Chiron, (10199) Chariklo, (38628) Huya, (50000) Quaoar, (55637) 2002\,UX25, (84522) 2002\,TC302, (90482) Orcus, (120347) Salacia, (136108) Haumea \\
    Kiss et al. (2013) & {\bf [K13]} & 2012 DR30 \\
    Lellouch et al. (2010) & {\bf [L10]} & (136108) Haumea \\
    Lellouch et al. (2013) & {\bf [L13]} & (20000) Varuna, (55636)2002 TX300, (120348) 2004 TY364, (15820) 1994 TB, (28978) Ixion (33340) 1998 VG44, (26308) 1998 SM165, (26375) 1999 DE9, (119979) 2002 WC19, (44594) 1999 OX3, (48639) 1995 TL8 \\
    Lellouch et al. (2016) & {\bf [L16]} & Pluto+Charon \\
    Lim et al. (2010) & {\bf [LIM10]} & (136472) Makemake, (90482) Orcus \\ 
    Mommert et al. (2012) & {\bf [MM12]} & 18 plutinos \\ 
    M\"uller et al. (2010) & {\bf [M10]} &  (208996)~2003~AZ84, (126154)~2001~YH140, (79360)~Sila-Nunam, (82075)~2000~YW134, (42355) Typhon, 2006~SX368, (145480)~2005~TB190  \\ 
    M\"uller et al. (2019) & {\bf [M19A]} & (136081) Haumea \\
    P\'al et al (2012) & {\bf [P12]} & (90377) Sedna, 2010\,EK\,139\\ 
    P\'al et al. (2015) & {\bf [P15]} & 2013~AZ\,60 \\ 
    Santos-Sanz et al. (2012) & {\bf [SS12]} & 15 scattered disk and detached objects \\ 
    Santos-Sanz et al. (2017) & {\bf [SS17]} & (84922) 2003 VS\,2, (208996) 2003 AZ\,84\\ 
    Vilenius et al. (2012) & {\bf [V12]} & 19 classical transneptunian objects \\ 
    Vilenius et al. (2014) & {\bf [V14]} & 18 classical transneptunian objects \\
    Vilenius et al. (2018) & {\bf [V18]} & 1995~SM\,55, 2005~RR\,43, 2003~UZ\,117, 2003~OP\,32, 2002~TX\,300, 1996~TO\,66, 1999~CD\,158, 1999\,KR\,16 \\
    \hline
    \end{tabular}
    \caption{Resources of Herschel Space Observatory measurements}
    \label{tab:herschelrefs}
\end{table*}
\normalsize
\paragraph{WISE:} The WISE mission is described in \cite{Wright2010}. Data products are summarized in the Explanatory Supplement to the AllWISE Data Release Products \citep{wiseexpsup}. 

 The WISE Moving Object Pipeline Subsystem (WMOPS) reported all detections of Solar System small bodies to the IAU Minor Planet Center (MPC) for confirmation, whereas the computed in-band magnitudes were collected in the IRSA/IPAC archive, namely in the Level 1b catalogues. To retrieve these magnitudes, we queried the IPAC archive using a 1~arcsec cone search radius around the MPC-reported tracklets, which are all labelled 'C51' by the MPC. In this way, we avoid using false detections that may have been included in the IPAC archive \citep{Mainzer2011}.

Since we are only interested in flux densities collected during the fully-cryogenic phase of the mission, we queried the WISE All-Sky Database. The in-band magnitudes (m) were converted to in-band flux densities ($\langle f \rangle$) as:
\begin{equation}
\langle f \rangle = \langle f_0 \rangle 10^{-0.4m}
, \end{equation}
where $\langle f_0 \rangle$ is the zero-magnitude isophotal flux density of Vega for each band, as reported in \cite{Wright2010}. By definition, $\langle f_0 \rangle$ does not require a colour correction. From the tabulated magnitude error bar $\Delta m$, the corresponding error bar of the in-bad flux is given by: 
\begin{equation}
    \Delta f = 0.40 \log_{10} \langle f \rangle \Delta m 
.\end{equation}

To correct for a discrepancy between red and blue calibrators observed after launch, \citealt{Wright2010} suggest shifting the W3 and W4 isophotal wavelengths and correcting the isophotal flux densities accordingly. Thus, we took 11.10 and 22.64$\mu$m and 31.368, and 7.952 Jy, respectively (more details in \cite{Masiero2011}). Flux densities obtained using this procedure are referred to as [{\bf WISEASD}] in the respective field of our IRDB. In this release of the database, we included only those WISE measurements where there is also AKARI data for that object.

\section{Auxiliary data \label{sect:aux}}

In our database, in addition to the basic data available in the resources listed above (measured (in-band) flux density and its uncertainty, date, instrument and filter, etc.), each measurement is supplemented with additional (auxiliary) data, either with data from external resources (mainly JPL/Horizons, Sect.~\ref{sect:jpl}), as well as values calculated from other fields of the database. The latter include the calculation of the monochromatic flux density at a pre-defined reference wavelengths (colour correction), conversion of calendar date to Julian date  (JD, with or without correction for light-travel time), and adding absolute flux density errors (Sect.~\ref{sect:calcval}).
The procedures to obtain these auxiliary data are summarized below. 

\subsection{JPL Horizons data} \label{sect:jpl}

Our database uses data obtained from NASA's JPL-Horizons service. These data are stored in the database and available directly in the IRDB and, in some cases, also used for further calculations. We query the following parameters from JPL-Horizons: 

\begin{itemize}
\item Orbital elements: semi-major axis, a (AU); eccentricity, e; inclination w.r.t XY-plane, i (degrees) (XY-plane: plane of the Earth's orbit at the reference epoch\footnote{Obliquity of 84381.448 arcseconds with respect to the ICRF equator (IAU76)}; longitude of ascending node, $\Omega$ (degrees); argument of perifocus, $\varpi$ (degrees);  mean anomaly, M (degrees).
\item Parameters related to size and albedo: absolute magnitude, H (mag); slope parameter G;  object's effective radius [km];  object's V-band geometric albedo. 

\item Apparent position: apparent right ascension (RA) and declination (DEC) at the time of the observation, ICRF/J2000 inertial reference frame, compensated for down-leg light-time [deg/deg];  rate of change of target center apparent R.A. and DEC (airless). We note that dRA/dt is multiplied by the cosine of the declination [both in arcsec\,hour$^{-1}$]. 
\item Target brightness: asteroid's approximate apparent visual magnitude and surface brightness, 
\begin{equation}
AP_{mag} = H + 5\log_{10}(\Delta) + 5\log_{10}(r) - 2.5\log_{10}((1-G)\Phi_1 + G\Phi_2) ,\end{equation}
where $H$ is the absolute brightness, $r$ the heliocentric distance of the target, $\Delta$ is the observer distance, and $G$, $\Phi_1$ and $\Phi_2$ are the slope parameter and the two base functions in \citet{Bowell} [mag, mag\,arcsec$^{-2}$]. 

\item Heliocentric distance: heliocentric range ($r$, light-time corrected) and range-rate ("rdot") of the target center [AU and km\,s$^{-1}$]. In addition, the one-way down-leg light-time from target center to observer is also retrieved [sec]. 

\item Sun-Observer-Target angle: target's apparent solar elongation seen from the observer location at print-time [degrees]. '/r' flag indicating the target's apparent position relative to
the Sun in the observer's sky ('/T' for trailing, '/L' for leading). 

\item Ecliptic coordinates of the target: Observer-centered Earth J2000.0 ecliptic longitude and latitude of the target center's apparent position, adjusted for light-time, the gravitational
deflection of light and stellar aberration [deg/deg]. 

\item X, Y, Z Cartesian coordinates of the target body and observer/Earth at the time of observation in the J2000.0 reference frame defined in Archinal et al. (2011) [AU].  
\end{itemize}

Data from JPL Horizons are obtained by four functions in the main python script. 
These functions are:
\begin{itemize}
\item {\sl get\_jplh}: this function obtains the ephemeris data from JPL Horizons, such as RA, DEC, r, delta, etc.
\item {\sl get\_jplhelements}: the orbital elements of the target body are retrieved with this function.
\item {\sl getvecs1 \& getvecs2}: these functions obtain the X,Y,Z position vectors from JPL with respect to the Sun and the observer.
\end{itemize}

The following parameters are also extracted from the .csv files obtained by these functions: H, JPL radius, albedo, and slope parameter. In those cases when JPL recognizes the target as a comet and not an asteroid, T-mag (Comet's approximate apparent visual total magnitude), M1 (Total absolute magnitude) parameters are extracted and used. Also, a comment is written into to the database in this case.

The functions use astropy.table from Astropy \citep{astropy2013,astropy2018} and urllib3\footnote{https://pypi.org/project/urllib3/}.

\subsection{Calculated values \label{sect:calcval}}

\paragraph{Obtaining monochromatic flux density (colour correction):}
For most of the instruments or filters included in the IRDB, colour correction is calculated using the relative response profiles of the specific filters and assuming an estimated effective temperature (T$_{\mathrm{eff}}$) for the target, which is calculated as:  
\begin{equation}
T_{\mathrm{eff}} = {{393.6\,K}\over{\sqrt{r_h}}} (1-p_V \cdot q)^{1/4},
\label{eq:Teff}
\end{equation} 
where r$_h$ is the heliocentric distance (AU), p$_V$ is the V-band geometric albedo and $q$ is the standard Bowell (1989) phase integral: $q$\,=\,0.290+0.684G, where G is the slope parameter and using the solar constant of 1361.5 $W/m^{2}$ \citep{Kopp} and the Stefan-Boltzmann constant to get $T = 393.6\,K$ at 1 AU. We assumed an emissivity of unity. We use $T_{eff}$ to calculate a black body SED which we utilize during the colour correction. But in case of high-albedo objects this formula does not work well (see \citealt{Vilenius2018} and \citealt{Brucker2009}). If p$_V$ is unknown or uncertain, we use p$_V$\,=\,0.10 (see Sect. \ref{sect:albedos}). We note that the $p_V$ in Eq. (\ref{eq:Teff}) is obtained from the tabulated $H$ and $D$ values from the Horizons JPL service. Arguably, there are other sources of diameters and H values that could be used to calculate $p_V$ but the colour correction values are not extremely sensitive to large variations of $p_V$ (see Sec. \ref{sect:albedos} for further discussion of this point). We include these diameters, albedos, and H-G values in the database for the sake of reproducibility of the monochromatic flux densities, but for statistical analyses with these quantities we recommend using or at least examining other sources (e.g. \citealt{Delbo2017}, \url{https://www-n.oca.eu/delbo/astphys/astphys.html}, \citealt{Oszkiewicz2011}, \citealt{Veres2015}.
The colour corrected or monochromatic flux density is obtained as f$_\lambda$\,=\,f$_i$/K($\lambda$), where f$_i$ is the in-band flux density obtained directly from the measurements and K($\lambda$) is the colour correction factor, which is obtained using the spectral energy distribution of the source (flux density $F_{\nu}(\lambda)$) and the relative response of the detector/filter system ($R_{\nu}(\lambda)$) as:
\begin{equation}
    K(\lambda) = {{{1\over{F_{\nu}(\lambda_{c})}} \int F_{\nu}(\lambda) R_{\nu}(\lambda) d\lambda}\over{{1\over{F_{\nu}^{ref}(\lambda_{c})}} \int F_{\nu}^{ref}(\lambda) R_{\nu}(\lambda) d\lambda}}
, \end{equation}
where $\lambda_c$ is the central (reference) wavelength of the filter in the photometric system and  $F_{\nu}^{ref}(\lambda)$ is the reference spectral energy distribution of the photometric system (typically $\nu F_\nu$\,=\,const.).

The monochromatic flux density uncertainties are calculated as:
\begin{equation}
\delta f_\lambda = \sqrt{{{1}\over{K(\lambda)^2}} \big[ \Delta f_i^2 + (r_{abs} f_i)^2 \big ] + (r_{cc}f_\lambda)^2}
\label{eq:abserror}
,\end{equation} 
where $\Delta f_i$ is the in-band flux density uncertainty and $r_{abs}$ is the absolute calibration error, usually expressed as a fraction of the in-band flux (see below). The last term contains the flux density uncertainty due to the colour correction uncertainty characterised by $r_{cc}$, which is approximately proportional to the deviation of the actual value of the colour correction from unity.
In the present version, it is implemented in the following way: 
\begin{itemize}
    \item[-] if 0.95\,$\leq$\,$K(\lambda)$\,$\leq$\,1.05 then r$_{cc}$\,=\,0.01; 
    \item[-] if 0.90\,$\leq$\,$K(\lambda)$\,$\leq$\,0.95 then r$_{cc}$\,=\,0.02;
    \item[-] if $K(\lambda)$\,$\leq$\,0.90 or $K(\lambda)$\,$\geq$\,1.05 then r$_{cc}$\,=\,0.03. 
\end{itemize}
\paragraph{Procedure for adding the absolute calibration error:}
Absolute calibration error is calculated as described above in Eq.~\ref{eq:abserror}. The $r_{abs}$ factor is instrument- or filter-dependent and it is determined during the flux calibration of the instrument and is described in instrument specific calibration papers. 
These $r_{abs}$ values are following:
\begin{itemize}
    \item IRAS: 10\%, 10\%, 15\% and 15\% at 12, 25, 60 and 100\,$\mu$m  \citep{Tedesco2002}
    \item Herschel/PACS: 5\%, 5\% and 7\% at 70, 100 and 160\,$\mu$m  (Balog et al., 2014) 
    \item WISE: 5\% in all bands \citep{Wright2010,Rozitis2014,Hanus2015}
    \item MSX \footnote{We took the absolute calibration uncertainties from Table 1 except for band A, for which we used a more conservative 5\% absolute calibration error bar.}: 5\%, 5\%, 6\% and 6\% at 8.28, 12.13, 14.65 and 21.34\,$\mu$m \citep{Egan2003}
    \item AKARI: 5\% in all bands \citep{Usui2011,Vali2018}
\end{itemize} 

\section{Database and access \label{sect:database}}

\subsection{Main database file and web interface}

From the collected data (flux densities, observational meta data, and auxiliary data, see above) a PostgreSQL table is created that is the essentially the SBNAF Infrared Database. The database is accessible through a web interface, available at https://ird.konkoly.hu/ (at Konkoly Observatory). An example is shown below, presenting the query screen (Fig. \ref{fig:ceresquery1}) and the resulting output screen (Fig. \ref{fig:ceresquery2}) for some selected IRAS and Akari observations of 1~Ceres. 
\begin{figure*}[ht!]
  \centering
  \includegraphics[width=1.0\linewidth]{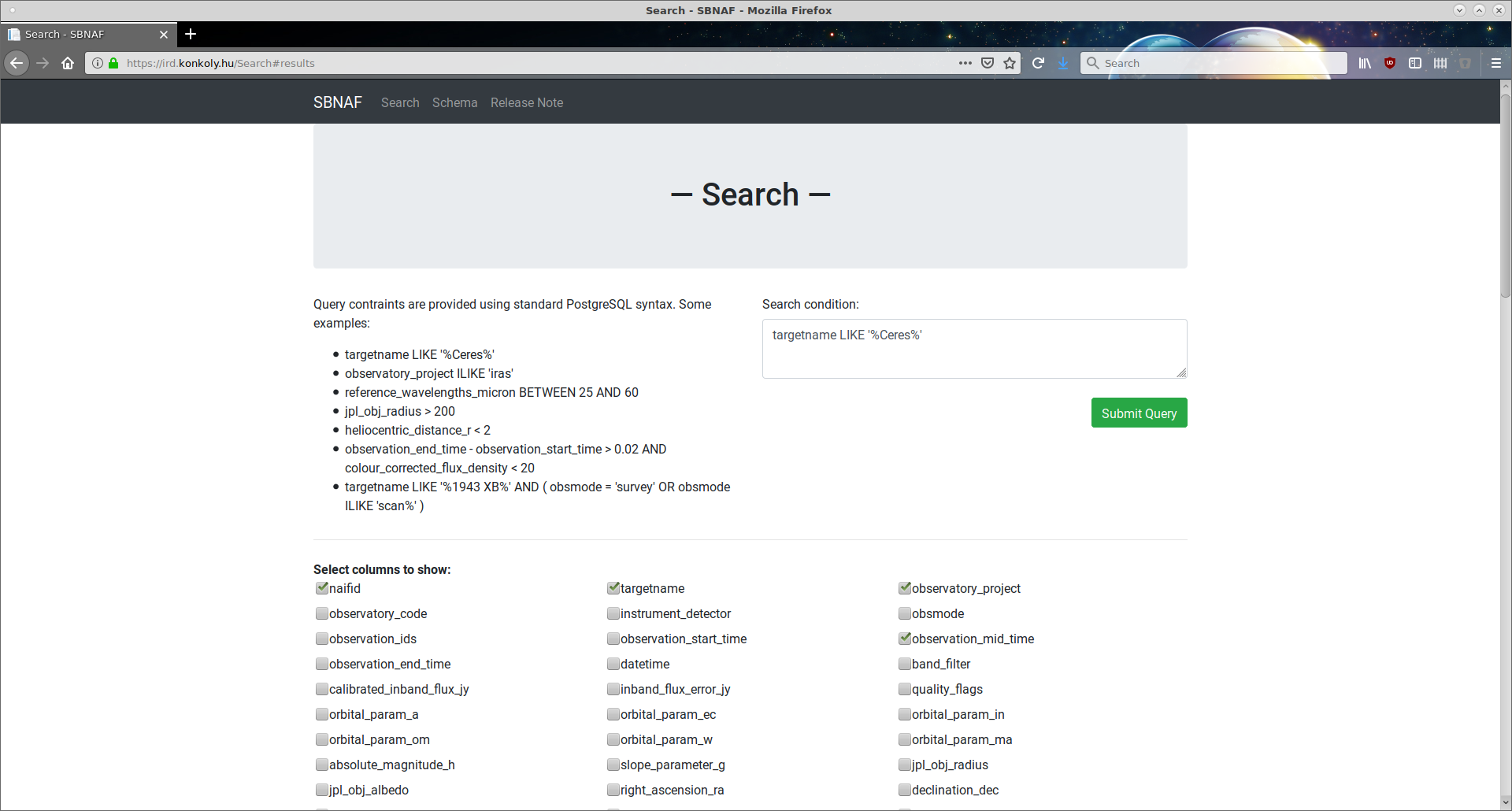}
  \caption{Query of 1~Ceres, with some examples on the left.  }\label{fig:ceresquery1}
\end{figure*}

\begin{figure*}[ht!]
  \centering
  \includegraphics[width=1.0\linewidth]{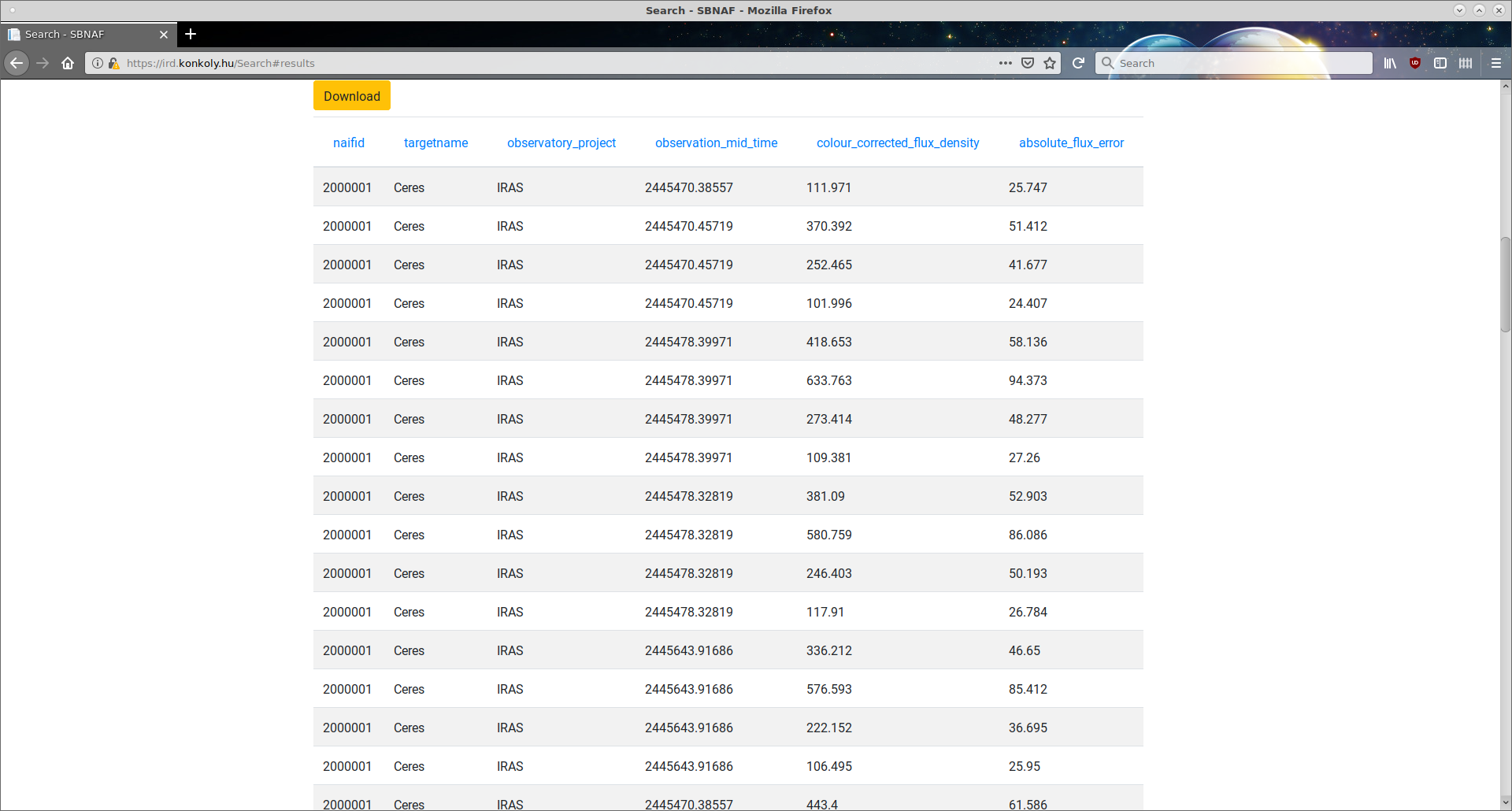}
  \caption{Results listing infrared observation of 1~Ceres and displaying the default output selection: 
  'naifid', 'targetname', 'observatory\_project', 'observation\_mid\_time', 'colour\_corrected\_flux\_density', and 'absolute\_flux\_error'.
  }\label{fig:ceresquery2}
\end{figure*}


The SBNAF IRDB webpage is running on a dedicated linux machine. The webserver is NGINX and the webpage itself is a ASP.NET Core Web Application, which uses the ASP.NET Core 2.0 Framework. This web application communicates with the webserver, with the users through the webpage and with the database itself (PostgreSQL table). The web content is generated on the server, and the users can only see the static HTML page in their web browser at the end. The user has the option to download the output page as a .csv file (using the 'Download' button). 

The example above was generated by the following query: {\bf targetname LIKE \textquotesingle\%Ceres\%\textquotesingle and (observatory\_project LIKE \textquotesingle IRAS\textquotesingle or observatory\_project ILIKE \textquotesingle Akari\textquotesingle)}, requiring exact, case sensitive matching with 'LIKE' on 'targetname' (='Ceres') and selecting observatories 'IRAS' (exact name again with 'LIKE') and 'Akari' (selecting with 'ILIKE', i.e. case-insensitive). In this case, the output gives the default selection: 'naifid', 'targetname', 'observatory\_project', 'observation\_mid\_time', 'colour\_corrected\_flux\_density', and 'absolute\_flux\_error'. 

Numeric data types (FLOAT, DOUBLE) can be queried as follows:
{\bf observation\_start\_time BETWEEN 2453869 AND 2454000}: selects the objects where the observation time started between 2453869 and 2454000 (JD); 
{\bf observation\_start\_time $>$ 2453869 AND observation\_start\_time $<$ 2454000}: returns the same results as the previous query. {\bf naifid = 2000001} : returns all targets with 'naifid' exactly equal to 2000001. 

For string type data:
{\bf targetname ILIKE \textquotesingle\%ceres\%\textquotesingle}: selects the targets with the substring "ceres" in it, case insensitive;
{\bf observatory\_project = \textquotesingle IRAS\textquotesingle}: returns all observations where the observatory\_project name matches 'IRAS' exactly. 

The query "{\bf comments\_remarks ILIKE \textquotesingle\%comet\%\textquotesingle}" lists measurements of comets (currently the Akari observations of the comet P/2006 HR30 (Siding Spring) in the D2.5 version of the database). 

Additional examples are given on the starting page of the web interface. The summary of the database fields is given in the Appendix.

\subsection{Alternative access}
For those who want a more scriptable form of access to the IRDB, we will upload the database into the VizieR catalog service \citep{vizier}. This database is widely used by the community and it provides an interface to search trough the different databases, and it can be queried trough several tools, like Astropy's astroquery.vizier interface or Topcat. In case of a major update we will upload a new data release to VizieR.

Also, we began the integration of the IRDB to the VESPA VO service\footnote{\url{https://voparis-wiki.obspm.fr/display/VES/EPN-TAP+Services}} \citep{Erard2018}. In this way our database can be accessed via EPN-TAP protocol, which is well known in the Virtual Observatory community. VESPA also provides a great tool for searching trough the different data services and with its help, the IRDB can become more accessible to researchers working on topics related to small body thermal emission. 
Because the VEASPA service will use the same database as our webpage, the data entered there will be always up to date. 

The whole database can be downloaded directly from our webpage in csv format. The datafile contains a header and exactly the same data that can be accessed via the SQL query form. The size of the raw csv file is $\sim$125 MB in the current release.

\section{Albedos and colour corrections of transneptunian objects} \label{sect:albedos}

As described in Sect~\ref{sect:jpl}, albedo information is taken by default from the NASA/Ho\-rizons service. In case these albedo values are not known, we assumed an intermediate value of p$_V$\,=\,0.10 to calculate the colour correction, which is a reasonable approximation for objects with albedos p$_V$\,$\lesssim$\,0.30. We  investigated how different the colour correction factors obtained using the Horizons albedo values are from those obtained using albedos calculated for specific objects based on radiometry in previous studies \citep[e.g.][]{Mommert2012,Santos2012,Vilenius2012,Vilenius2014}. This comparison was performed for Centaurs and transneptunian objects, as in these cases significant deviations are expected from the general, low albedo values.

\begin{figure*}[ht!]
  \centering
  \includegraphics[width=1.0\linewidth]{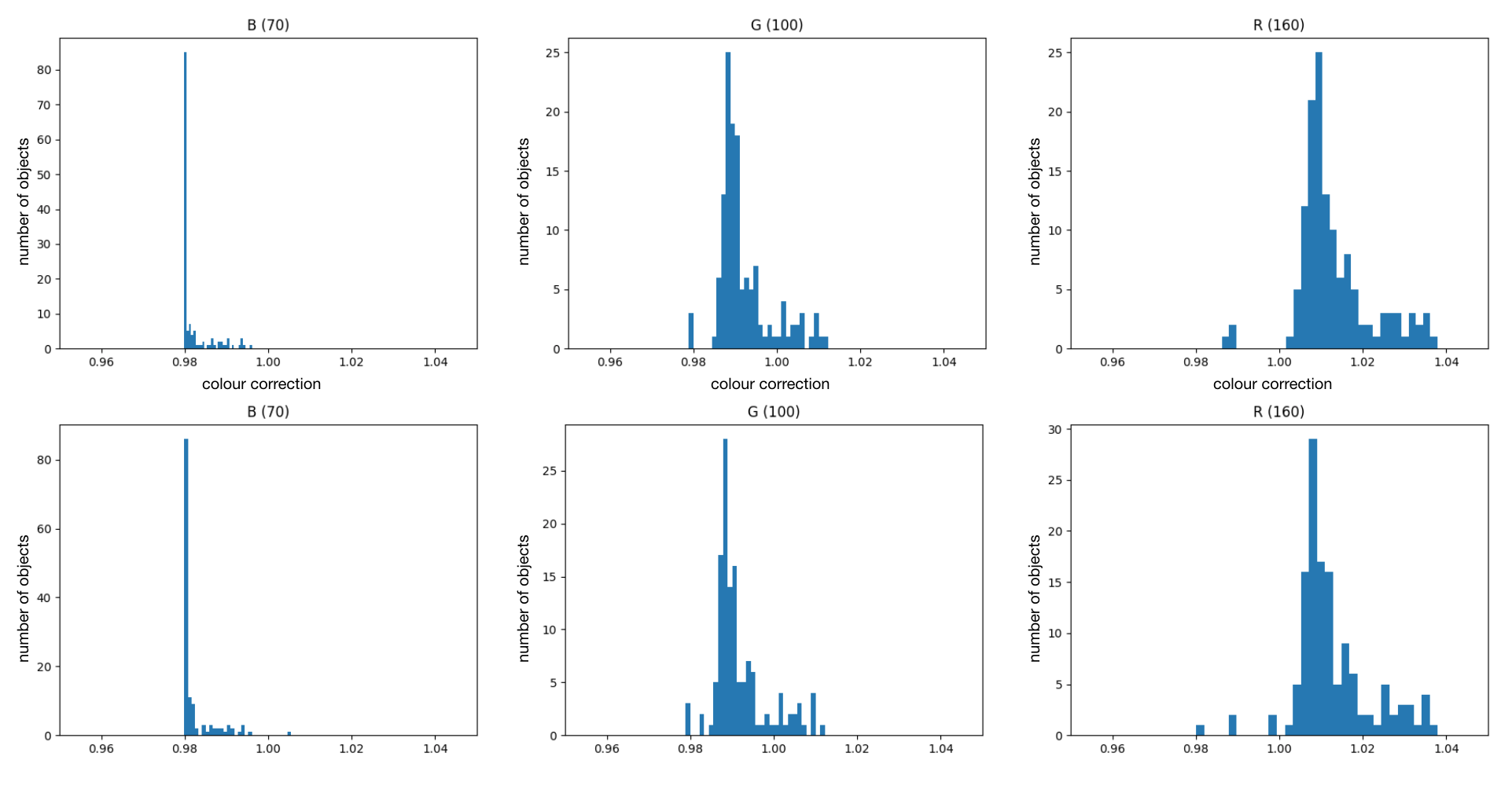}
  \caption{Comparison of the colour correction factors obtained using the 'default' albedo data from NASA/Horizons (upper row), and the albedos obtained from specific publications and/or processing. B(70), G(100) and R(160) mark the 70, 100 and 160\,$\mu$m filters of the PACS camera. 
  }\label{fig:cc}
\end{figure*}


Fig.~\ref{fig:cc} presents the colour correction factors obtained using the albedo values from NASA/Horizons (top row), as well as those obtained from specific publications (bottom row). In general, the differences in the colour correction factors are negligible (mainly because the corrections factors typically close to 1 in the surface temperature range for the Herschel/PACS filters). However, for some individual targets (e.g those with large heliocentric distance and relatively high albedo) the colour correction factors may differ by 5--10\%. Therefore, in the updated version of the infrared database, we used the object-specific values whenever they are available.  

\section{Thermal inertias of asteroid populations\label{sect:asterpoidthermalinertia}}

In this section, we provide an example of how the database can be used to study scientific questions. 
Namely, we show how to explore the dependence of thermal inertia ($\Gamma$) with temperature ($T$) or, equivalently, heliocentric distance ($r$), and how to compare populations of asteroids with different sizes. 
These quantities are relevant, for example,  for studies of dynamical evolution where the Yarkovsky effect plays a role since it is influenced among other factors by $\Gamma$ \citep[see e.g.][]{Bottke2006}.

Thermal inertia is defined as $\Gamma = \sqrt{\rho C\kappa}$, where $\rho$ is the density, $C$ the specific heat capacity, and $\kappa$ the thermal conductivity of the material. Its SI unit is J\,m$^{-2}$s$^{-1/2}$K$^{-1}$. Most thermo-physical models of asteroids used so far to constrain $\Gamma$ have assumed it is constant even though the heat conductivity is itself a function of temperature. For this reason, thermal inertias determined from observations taken at different heliocentric distances cannot be compared directly and must be normalised to an agreed heliocentric distance (see, e.g. \citealt{Delbo2015} for a review). 
Assuming that heat is transported through the regolith mainly by radiative conduction within the spaces between the grains \citep{Jakosky1986,Kuhrt1989}, we have $\kappa \propto T^3$ and hence $\Gamma \propto r^{-3/4}$ (see also the discussion by \citealt{Delbo2007}). We can then write
\begin{equation}\label{eq:gamma}
  \Gamma(r) = \Gamma_0 r^{\alpha},
\end{equation}
where $\Gamma_0$ is the value at 1 au, $r$ is expressed in au, and $\alpha=-3/4$ is the classically-assumed radiative conductivity exponent.

\citet{Rozitis2018} studied three eccentric near-Earth asteroids observed by WISE more than once at a wide range of $r$ and showed that $-3/4$ may not be appropriate in general. 
Instead, they found that each one of their targets required different exponents to fit their $\Gamma$-vs.-$r$ plots, namely $\alpha = -2.2, -1.5,$ and $-0.92$, although only the error bars of the first case, (1036) Ganymed, were incompatible with the `classical' $\alpha = -3/4$. 
This effect requires further studies. Below, we propose ways to use our database to examine how $\alpha$ affects our interpretation of the thermal IR observations. 

\begin{figure}
  \centering
  \includegraphics[width=\linewidth]{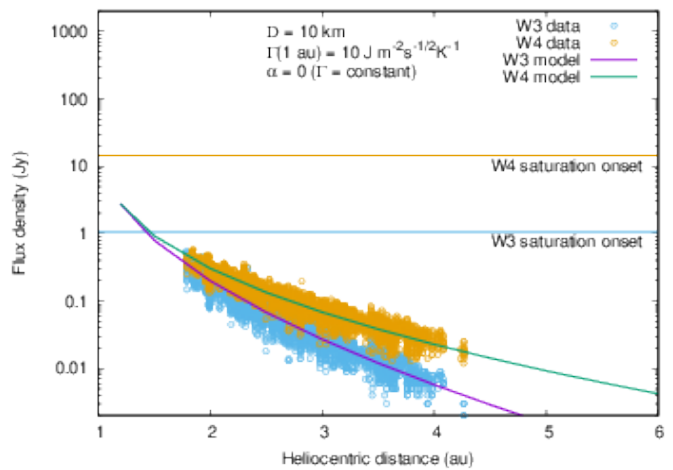}
  \caption{WISE W3 and W4 colour-corrected fluxes of 10-km asteroids versus heliocentric distance (blue and ochre empty circles). 
    The purple and green lines are the corresponding model fluxes with $\Gamma_0=10$\,J\,m$^{-2}$s$^{-1/2}$K$^{-1}$ and $\alpha=0$. 
    For reference, since we have excluded partially saturated fluxes from the analysis (cf. \ref{fig:flux_a_220_75km}, left panel), we also indicate the fluxes at which saturation occurs in each band \citep{Cutri2012}. 
  }\label{fig:flux_a_cte}
\end{figure}

With our database, we can readily select the thermally dominated WISE W3 and W4 fluxes of a group of asteroids within a given size and albedo range and compare how models with different pairs of values of $\Gamma$ and $\alpha$ fit the observations. First, we took 10-km, low-albedo asteroids. To account for uncertainties, we queried the database for all W3 and W4 colour-corrected fluxes of objects in the size range $9\,\mathrm{km} \leq D \leq 11\,\mathrm{km}$ with visible geometric albedos $p_V \leq 0.12$. 
This sample includes $\sim$4000 fluxes per band, corresponding to about 300 asteroids, the vast majority of which are main-belt asteroids. 
Figure~\ref{fig:flux_a_cte} shows these fluxes plotted against heliocentric distance at the epoch of observations. 
For reference, we also show the fluxes at which the onset of saturation occurs in each band with horizontal lines of the corresponding colour. 

Next, we used the thermo-physical model of \citet{Delbo2007} under the Lagerros approximation \citep{Lagerros1996I,Lagerros1998} to calculate the W3 and W4 fluxes at various heliocentric distances ($r = 1.25, 1.5, 2.0, 2.5, \ldots 6.0$ au) for a rotating sphere\footnote{
  More specifically, we averaged the fluxes of a prograde sphere and a retrograde sphere with pole ecliptic latitudes of $\pm$90 degrees, respectively. In addition, to reproduce WISE's observation geometry, the model fluxes are computed for an observer-sun-asteroid configuration in quadrature assuming a circular orbit on the ecliptic.
}
with rotation period $P = 6$\,h, 

Bond albedo $A=0.03$, surface roughness rms = 0.6 (the default value for main belt asteroids chosen by \citealt{Mueller1998,Mueller2002}), and different thermal inertias at 1 au. We sampled $\Gamma_0 =$\,10, 30, 50, 100, 150, 200 J\,m$^{-2}$s$^{-1/2}$K$^{-1}$ and $\alpha =$\,0, $-$0.75 and $-$2.20, and used Eq.~\ref{eq:gamma} to compute the corresponding $\Gamma(r)$ for each case. 

\begin{figure}
  \centering
  \includegraphics[width=\linewidth]{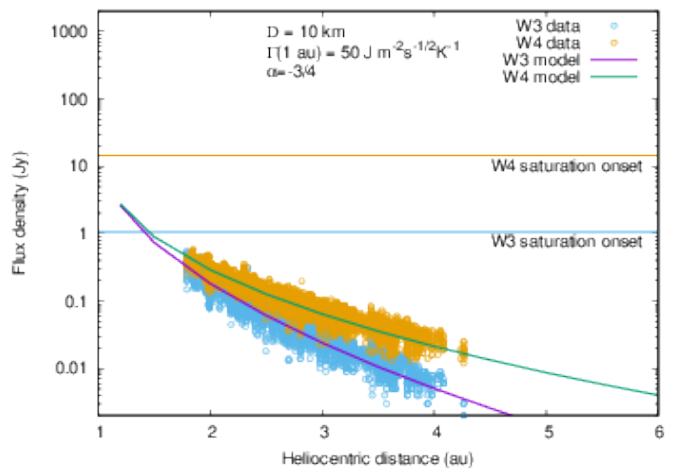}
  \caption{Same as Fig.~\ref{fig:flux_a_cte} but with $\alpha=-3/4$ and $\Gamma_0=50$\,J\,m$^{-2}$s$^{-1/2}$K$^{-1}$ 
  }\label{fig:flux_a_075} 
\end{figure}

\begin{figure}
  \centering
  \includegraphics[width=\linewidth]{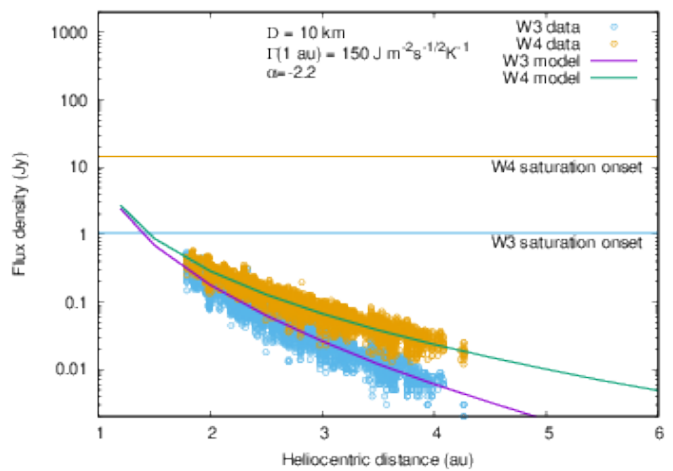}
  \caption{Same as Fig.~\ref{fig:flux_a_cte} but with $\alpha=-2.2$ and $\Gamma_0=150$\,J\,m$^{-2}$s$^{-1/2}$K$^{-1}$ 
  }\label{fig:flux_a_220} 
\end{figure}

Figures~\ref{fig:flux_a_cte} to \ref{fig:flux_a_220} show the W3 and W4 data and the best fitting models corresponding to $\alpha = 0, -0.75$, and $-2.20$. The best $\Gamma_0$\, in each case increases with $\alpha$, namely 10, 50 and 150 J\,m$^{-2}$s$^{-1/2}$K$^{-1}$. 
  These models appear to reproduce the data trend reasonably well considering the numerous approximations that we made.
  To establish a quantitative comparison, we averaged the data within 0.5 au-wide heliocentric distance bins centred at 2.0, 2.5, \ldots 4.5 au, and used these averages ($f_{\mathrm{bin}}$) and standard deviations ($\sigma_{\mathrm{bin}}$) to calculate the $\chi^2$ for each model ($F_{\mathrm{bin}}$)
  as $\chi^2 = \sum_{\mathrm{bins}}\left[\left(F_{\mathrm{bin}}-f_{\mathrm{bin}}\right)/\sigma_{\mathrm{bin}}\right]^2$. 
  Considering two degrees of freedom, $\alpha$ and $\Gamma_0$, all three cases produced reduced $\chi^2$ between $\sim$4.5 and 5 indicating that the data are fit at ~2.2 sigma only in average.
  Hence, we cannot simultaneously optimise $\alpha$ and $\Gamma_0$ with this approach given the current lack of `ground-truth' information for objects in this size range.

  Nonetheless, it is still worth putting these models in the context of other populations. For example, if $\alpha = 0$ was to be confirmed, we would conclude that the typical $\Gamma_0$'s of 10-km asteroids would be a factor of 5 lower than that of the Moon's ($\sim$50 J\,m$^{-2}$s$^{-1/2}$K$^{-1}$). This would be unexpected because so far such low $\Gamma$s have only been found among the largest asteroids \citep[e.g.][]{Mueller2002,Mueller2006,ORourke2012,Delbo2009,Hanus2018}, but we cannot rule the possibility out from this approach alone.
  Among the models with $\alpha=-3/4$, the best $\Gamma_0$ is very close to the lunar value and the average $\Gamma_0$ of the nine 10-km objects in Table A.3 of \citet{Hanus2018}, namely $\sim$58 J\,m$^{-2}$s$^{-1/2}$K$^{-1}$. 
  However, this does not confirm $\alpha=-3/4$ as the best value for 10-km either. Based on previous modelling reviewed by \cite{Delbo2015} ;
  see their right panel of Fig. 9; we would actually expect higher $\Gamma_0$s for the `classical' $\alpha$ value, specifically in the $\sim 300$ J\,m$^{-2}$s$^{-1/2}$K$^{-1}$ (the role of size is further discussed below).  
  A larger sample of 10-km asteroids with accurate shapes and modelled $\Gamma$s would be needed to establish a more conclusive result. 
  We would also need many more high-quality shape models of highly eccentric objects to enable additional studies like \citet{Rozitis2018} in order to constrain $\alpha$, given its crucial impact on the $\Gamma_0$ values and ultimately on their physical interpretation, especially in the light of the results from the space missions. 

\begin{figure*}
  \centering
  \includegraphics[width=0.45\linewidth]{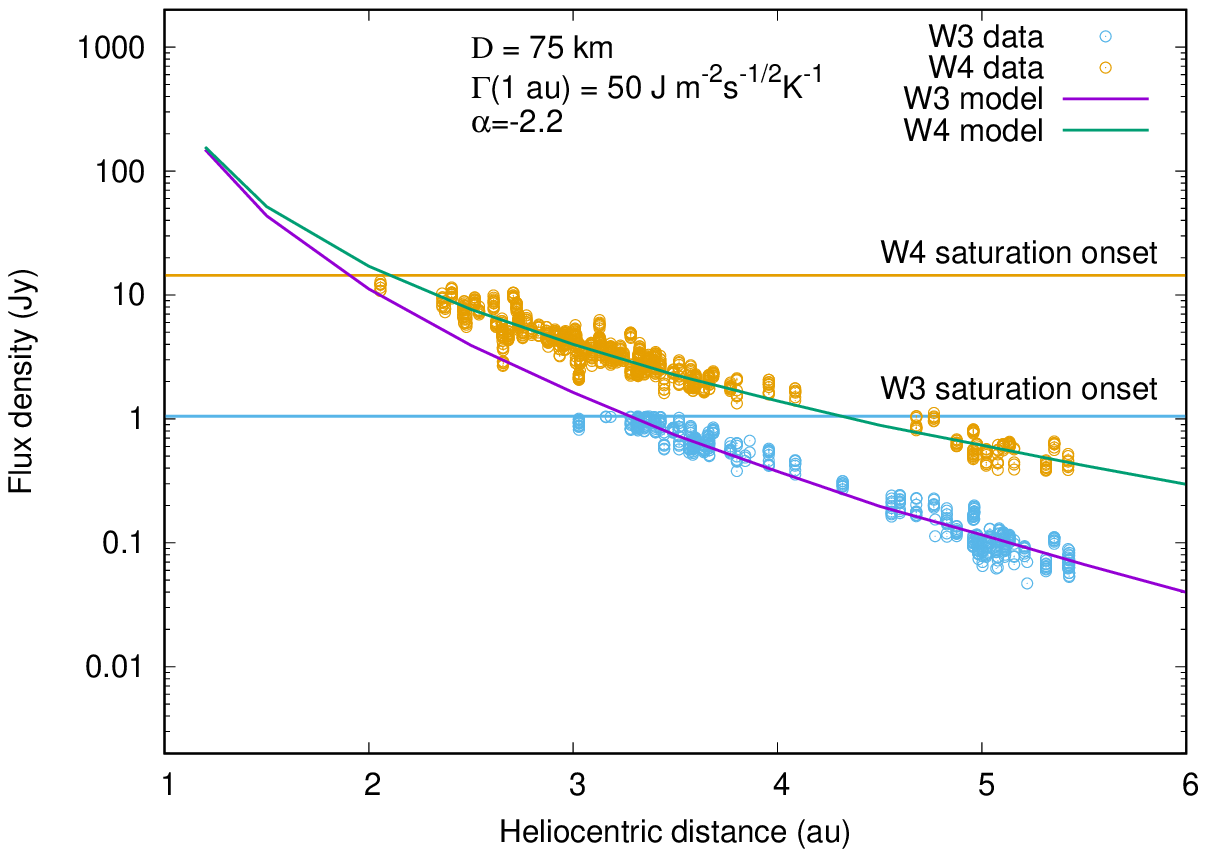}
  \includegraphics[width=0.45\linewidth]{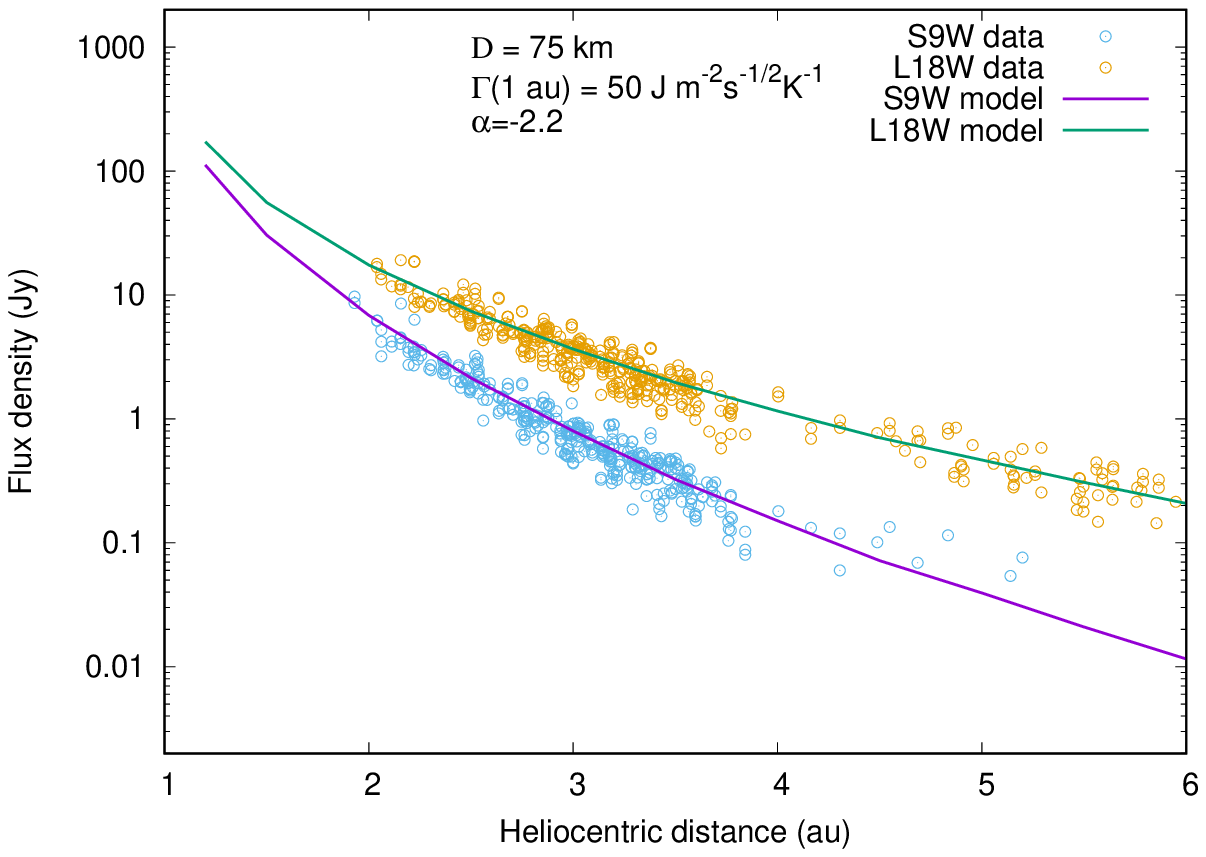}
  \caption{Left: the onset of partial saturation limits the usefulness of the W3 data for the 75-km population. 
    The gap between the main belt and the Hildas/Cybeles/Jupiter Trojans is apparent. 
    Right: The same plot but using AKARI data.
    In this case there were no saturation limits but there were fewer 18-$\mu$m data at $r > 4$\,au. 
  }\label{fig:flux_a_220_75km}
\end{figure*}

The approach illustrated here does enable the comparison of thermal inertias of differently sized populations if we keep all other parameters equal. 
  For instance, we have about 2500 observations of the $\sim$100 objects with sizes of 75 km that are featured both in the WISE and the AKARI catalogues (see Sec.~\ref{sect:keyref}). 
  With $\alpha=-2.20$, we require a lower $\Gamma_0$ to fit 75-km asteroid data (50 J\,m$^{-2}$s$^{-1/2}$K$^{-1}$) than 10-km asteroid data (150 J\,m$^{-2}$s$^{-1/2}$K$^{-1}$). 
  The W3 and W4 (left panel) and S9W and L18W (right panel) data and corresponding model fluxes for 75-km asteroids are shown in Fig.~\ref{fig:flux_a_220_75km}.
  Although there are fewer objects in this size range and removing partially saturated data introduces an artefact in the W3 data trend, the conclusion that smaller asteroids have higher thermal inertias is robustly supported, at least when comparing the 10 to 75-km diameter range. 


  Finally, some of the effects not accounted for by our simplifying assumptions could be added into the modelling in future work, such as, a large set of shapes with different elongations or degrees of irregularity and spin pole orientations. With these, and a finer and more complete sampling of $\Gamma$, $\alpha$, and $P$, it might be possible to obtain a model with a high `goodness of fit' \citep[see e.g.][]{Press1986}.
  We have not considered the completeness of the samples and we emphasise that  the database does not currently contain the full WISE catalogue, but only the data pertaining to those objects that are featured in the AKARI catalogue. 
The AKARI catalogue is complete in the Main Belt down to sizes $~$20 km \citep{Usui2013,Usui2014}, but contains fewer Hildas/Cybeles/Jupiter Trojans in the L18W band. 
Also, in future works, we will incorporate the full WISE catalogue into the database to have more NEAs and members of populations beyond the main belt. 

\section{Future updates of the SBNAF Infrared Database}\label{sect:updates}
\label{updates}

\subsection{Planned updates}

A full assessment of {\it all} Herschel/PACS main belt asteroid observations were performed in the framework of the Small Bodies: Near and Far Horizon 2020 project \citep{Muller2018}. Far-infrared flux densities obtained from these measurements will be made public with the next release of the database. Infrared data from serendipitous observations of bright main belt asteroids with the Herschel/PACS and SPIRE photometer detectors are also planned to be added.

\subsection{Submitting data}

Researchers who have infrared measurements can submit their reduced data to the SBNAF Infrared Database by sending an email to the address {\it irdb@csfk.mta.hu}. After the submission the data will be processed as it was described in sections above, and these processed and auxiliary data will be uploaded to the database.
We request the input data in a CSV file (simple attachment of the cover email), with the following columns:

\begin{itemize}
    \item Number: if possible, asteroid number, e.g. 275809    \item Designation: provisional designation, e.g. 2001 QY297
    \item Observatory/project: name of the observatory or the space mission, e.g. Herschel, WISE, etc
    \item Observatory\_code: JPL/Horizons code of the observatory or spacecraft, e.g. 500@-486    \item Instrument\_detector: instrument of the observatory or spacecraft used in the specific measurement
    \item Obsmode: observation mode of the instrument, if available    \item Observation\_IDs: observation ID of the measurement, if available (e.g. AORKEY in the case of Spitzer, and OBSID in the case of Herschel measurements)
    \item Start: start time of the measurement in Julian date format   \item End: end time of the measurement in Julian date format   \item Band or filter: name of the filter or band used for the measurement
    \item Calibrated\_inband\_flux: in-band photometric flux density in [Jansky] units, with all photometric corrections applied, including aperture or encircled energy fraction corrections, but without colour correction    \item Inband\_flux\_error: uncertainty  of  the  in-band  photometric  flux  density ’calibrated\_inband\_flux’, with all direct photometric errors considered, but without errors related to the spectral energy distribution of the target (colour correction),and also without the consideration of the absolute photometric error of the instrument    \item Comments: comments regarding the measurement; e.g. the flux density is an upper limit, the flux is already colour corrected, or it is obtained from a combination of more than one measurements, etc\end{itemize}

\begin{acknowledgements}
This work has dedicatedly been supported by European Union’s Horizon 2020 Research and Innovation Programme, under Grant Agreement no 687378 ('Small Bodies: Near and Far'). The operation of the database is supported by the Hungarian Research, Development and Innovation Office (grant no. K-125015), and by the Hungarian Academy of Sciences (EUHUNKPT/2018). This research made use of Astropy,\footnote{http://www.astropy.org} a community-developed core Python package for Astronomy
\cite{astropy2013}, \cite{astropy2018}. 
We sincerely thank the referee for the thorough and detailed review.
\end{acknowledgements}



\clearpage
\newpage

\clearpage

\section*{Appendix}
\addcontentsline{toc}{section}{Appendix}
\section*{Summary of database fields}

Below we give a description of the output fields of the Infrared Database. The unit and the data type of the specific field are given in squared and regular brackets, respectively. 

\paragraph{naifid:} NASA's Navigation and Ancillary Information Facility\footnote{\url{https://naif.jpl.nasa.gov/pub/naif/toolkit\_docs/FORTRAN/req/naif\_ids.html}} solar system object code of the target (LONG)

\paragraph{targetname:} The name or designation of the asteroid. (STRING)

\paragraph{observatory\_project:} Name of the observatory/ space mission (STRING). The possible values are:
'IRAS': Infrared Astronomy Satellite; 
'AKARI': Akari Space Telescope; 
'MSX': Midcourse Space Experiment'; 
'WISE': Wide-field Infrared Survey Explorer; 
'HSO': Herschel Space Observatory. 

See Table\ref{table:instruments} for a list of instruments, filters and observatory codes used in this database. References for the listed instruments and flux density measurements can be found in Sect.~\ref{sect:keyref}. 

\paragraph{observatory\_code:} JPL/Horizons code of the observatory/spacecraft, see Table~\ref{table:instruments} for a list (STRING)

\paragraph{instrument\_detector:} Instrument of the observatory/spacecraft used in that specific measurement, see Table~\ref{table:instruments} for a list (STRING) (STRING)

\paragraph{obsmode:} Observation mode, also listed in Table~\ref{table:instruments}. For a descrition of the observing modes, see the respective references of the instruments in Sect.~\ref{sect:keyref}.  For instruments working in survey mode (like IRAS and MSX) with which no pointed observations were possible and data are taken from the survey data 'survey' in the 'obsmode' column indicates the default observing mode (STRING)

\paragraph{observation\_IDs:} Mission-specific identifier of the observation, e.g. {\it OBSID} for Herschel measurements and {\it AORKEY} for Spitzer observations [unitless] (STRING)

\paragraph{observation\_start\_time} Start time of the measurement      [Julian date] (DOUBLE)

\paragraph{observation\_mid\_time} Mid-time of the measurement [Julian date] (DOUBLE)

\paragraph{observation\_end\_time} End time of the measurement [Julian date] (DOUBLE)

\paragraph{datetime} Observation date in the format     'YYYY:MM:DD hh:mm:ss.sss', with YYYY: year; MM: month in string format (Jan, Feb, etc.), DD: day of the month; hh: hour of the day; mm: minutes; ss.sss: seconds with three-digit accuracy (STRING)

\paragraph{band\_filter} Name of the filter/band used for the specific observation (STRING)

\paragraph{calibrated\_inband\_flux\_Jy:} In-band photometric flux density in [Jansky] units, with all photometric corrections applied, including aperture/encircled energy fraction corrections, but without colour correction (DOUBLE). Sources of original data are listed in Sect.~\ref{sect:keyref}.

\paragraph{inband\_flux\_error\_Jy:} Uncertainty of the in-band photometric flux density 'calibrated\_inband\_flux\_Jy', with all direct photometric errors considered, but without errors related to the spectral energy distribution of the target (colour correction), and also without the consideration of the absolute photometric error of the instrument (DOUBLE). 

\paragraph{quality\_flags:}     (STRING)
\begin{itemize}
\item WISE\footnote{\url{http://wise2.ipac.caltech.edu/docs/release/allsky/expsup/index.html}}:
 \begin{itemize}
  \item  Contamination and confusion flags:
  \begin{itemize}
    \item P - Persistence. Source may be a spurious detection of (P).
    \item p - Persistence. Contaminated by (p) a short-term latent image left by a bright source.
    \item 0 (number zero) - Source is unaffected by known artifacts.
  \end{itemize}
  \item  Photometric quality flags:
    \begin{itemize}
      \item A - Source is detected in this band with a flux signal-to-noise ratio w\_snr\,$>$10.
      \item B - Source is detected in this band with a flux signal-to-noise ratio 3\,$<$\,w\_snr\,$<$\,10.
  \end{itemize}
 \end{itemize}
\end{itemize}
\paragraph{orbital\_param\_A:} semi-major axis of the target's orbit, as obtained from JPL/Horizons [AU] (DOUBLE)

\paragraph{orbital\_param\_EC:} eccentricity of the target's orbit, as obtained from JPL/Horizons [unitless (DOUBLE)

\paragraph{orbital\_param\_IN:} inclination of the target's orbit, as obtained from JPL/Horizons [deg] (DOUBLE)

\paragraph{orbital\_param\_OM:} longitude of the ascending node of the target's orbit, as obtained from JPL/Horizons [deg] (DOUBLE)

\paragraph{orbital\_param\_W:} argument of the periapsis of the target's orbit, as obtained from JPL/Horizons [deg]      (DOUBLE)

\paragraph{orbital\_param\_MA:} mean anomaly of the target's orbit, as obtained from JPL/Horizons [deg]                (DOUBLE)

\paragraph{absolute\_magnitude\_H}      
 absolute magnitde of the target, i.e. the visual magnitude an observer would record if the asteroid were placed 1\,AU away, and 1\,AU from the Sun and at a zero phase angle, as obtained from JPL/Horizons [mag]  (FLOAT)

\paragraph{slope\_parameter\_G} 'G' slope parameter of the target, describing the dependence of the apparent brightness on the phase angle (light scattering on the asteroid's surface); for more details, see Bowel (1989) (FLOAT). Note that the default slope parameter in the Horizons system is the canonical G\,=\,0.15, also used as a generally accepted value in applications when specific values are note available \citep[see e.g. ExploreNEO,][]{Trilling2010,Harris2011}.  

\paragraph{jpl\_obj\_radius:} estimated radius of the target as obtained from JPL Horizons [km] (FLOAT)

\paragraph{jpl\_obj\_albedo:} estimated V-band geometric albedo of the target as obtained from JPL Horizons   (FLOAT)

\paragraph{Right\_Ascension\_RA:} right ascension (J2000) of the target at observation mid-time, calculated from the orbit by JPL/Horizons [deg] (FLOAT)

\paragraph{Declination\_DEC:} declination (J2000) of the target at observation mid-time, calculated from the orbit by JPL/Horizons [deg]       (FLOAT)

\paragraph{RA\_rate:}  rate of change in right ascension [arcsec\,s$^{-1}$\,$\equiv$\,deg\,h$^{-1}$] (FLOAT)

\paragraph{DEC\_rate} rate of change in declination [arcsec\,s$^{-1}$\,$\equiv$\,deg\,h$^{-1}$]         (FLOAT)

\paragraph{apparent\_magnitude\_V} estimated apparent brightness of the target in V-band at observation mid-time, as obtained by JPL/Horizons [mag]
(FLOAT)

\paragraph{heliocentric\_distance\_r:} heliocentric distance of the target at observation mid-time, as obtained by JPL/Horizons  [AU] (DOUBLE)

\paragraph{obscentric\_distance\_delta:}        observer to target distance at observation mid-time, as obtained by JPL/Horizons  [AU]      (DOUBLE)

\paragraph{lighttime:}  elapsed time since light (observed at print-time) would have left or reflected off a point at the center of the target [sec] (FLOAT)

\paragraph{solar\_elongation\_elong:} target's apparent solar elongation seen from the observer location at print-time, in degrees (FLOAT)

\paragraph{before\_after\_opposition:} flag regarding the target's apparent position relative to the Sun in the observer's sky. '/T' indicates trailing, '/L' leading position with respect to the Sun   (STRING)

\paragraph{phase\_angle\_alpha:} Sun--Target--Observer angle at observation mid-time, as obtained by JPL/Horizons [deg] (FLOAT)

\paragraph{ObsEclLon/ObsEclLat:} observer-centered Earth ecliptic-of-date longitude and latitude of the
target center's apparent position, adjusted for light-time, the gravitational deflection of light and stellar aberration, in degrees, as obtained by JPL/Horizons [deg] (FLOAT)

\paragraph{target\_[X,Y,Z]@sun:} Sun-centered X, Y, Z Cartesian coordinates of the target body at observation mid-time, in the reference frame defined in Archinal et al. (2011) [AU].  (DOUBLE)

\paragraph{target\_[X,Y,Z]\_@observer:} observer-centered X, Y, Z Cartesian coordinates of the target body at observation mid-time, in the reference frame defined in Archinal et al. (2011) [AU].  (DOUBLE)

\paragraph{observer\_X\_@sun} Sun-centered X, Y, Z Cartesian coordinates of the observer at observation mid-time, in the reference frame defined in Archinal et al. (2011) [AU].  (DOUBLE) (DOUBLE)


\paragraph{reference\_wavelengths\_micron:} reference wavelength of the measuring filter in [$\mu$m] units        (FLOAT)

\paragraph{colour\_correction\_factor:} colour correction factor applied to obtain monochromatic flux density from in-band flux density [unitless]; see Sect.~\ref{sect:calcval} for details (FLOAT)

\paragraph{colour\_corrected\_flux\_density:} monochromatic flux density (colour corrected in-band flux density)  [Jy]; see Sect.~\ref{sect:calcval} for details (DOUBLE)

\paragraph{absolute\_flux\_error:} absolute uncertainty of the monochromatic flux density including the uncertainty of the absolute flux calibration [Jy]; see Sect.~\ref{sect:calcval} for details (DOUBLE)

\paragraph{comments\_remarks:}  comments regarding the quality of the measurement / information when an assumed value was applied in the calculations;e.g. indicating whether the target is (also) regarded as a comet in JPL/Horizons; non standard value of geometric albedo is also marked here when it is not taken from JPL/Horizons (STRING)

\paragraph{LTcorrected\_epoch:} The lighttime corrected epoch, calculated as \newline
$observation\_mid\_time - lighttime/3600./24.$ [day] (DOUBLE)

\paragraph{documents\_references:}      publications and resources which the photometric data were taken from (STRING). A list of codes can be found in Sect.~\ref{sect:keyref}.  

\paragraph{input\_table\_source:} name of the input file which the database was generated from (strictly for internal usage) (STRING)

\paragraph{data\_last\_modification:}  date when the record was last modified, in 'human readable' format: 'YYYY-MMM-DD hh:mm:ss' (STRING)

\paragraph{alt\_target\_name:} possible alternative names for the object, like asteroid number and provisional designation. (STRING)

\section*{Example of infrared database fields}

Below, we present the specification of the infrared database fields through an example (Akari measurement of (25143) Itokawa) as it was at the time of the production of this document. Changes may apply and the description should be taken from the latest version of the file {\sl AstIrDbTbl\_keys\_v$<$version\_number$>$.txt}.
\clearpage
\noindent\rule{\textwidth}{0.5pt}
\tiny
\begin{verbatim}
Parameter                         Type       Unit       ExampleValue                                Retrieved
                                                                                                     from JPL
-------------------------------------------------------------------------------------------------------------
01 naifid                         LONG       ---        2025143                                            No
02 targetname                     STRING     ---        Itokawa                                            No
03 observatory_project            STRING     ---        AKARI                                              No
04 observatory_code               STRING     ---        500@399                                            No
05 instrument_detector            STRING     ---        IRC-MIR-L                                          No

06 obsmode                        STRING     ---        IRC02                                              No
07 observation_IDs                STRING     ---                                                           No
08 observation_start_time         DOUBLE     days       2454308.04699                                      No
09 observation_mid_time           DOUBLE     days       2454308.04699                                      No
10 observation_end_time           DOUBLE     days       2454308.04699                                      No

11 datetime                       STRING     ---        2007-Jul-26 13:07:40.000                           No
12 band_filter                    STRING     ---        L15                                                No
13 calibrated_inband_flux_Jy      DOUBLE     Jy         0.02                                               No
14 inband_flux_error_Jy           DOUBLE     Jy         0.001                                              No
15 quality_flags                  STRING     ---                                                           No

16 orbital_param_A                DOUBLE     au         1.32404449                                        Yes
17 orbital_param_EC               DOUBLE     ---        0.28018482                                        Yes
18 orbital_param_IN               DOUBLE     deg        1.62206524                                        Yes
19 orbital_param_OM               DOUBLE     deg        69.09531692                                       Yes
20 orbital_param_W                DOUBLE     deg        162.77163687                                      Yes

21 orbital_param_MA               DOUBLE     deg        32.28481921                                       Yes
22 absolute_magnitude_H           FLOAT      mag        19.2                                              Yes
23 slope_parameter_G              FLOAT      ---        0.15                                              Yes
24 jpl_obj_radius                 FLOAT      km         0.165                                             Yes
25 jpl_obj_albedo                 FLOAT      ---        0.1                                               Yes

26 Right_Ascension_RA             FLOAT      deg        209.5522                                          Yes
27 Declination_DEC                FLOAT      deg        -16.11449                                         Yes
28 RA_rate                        FLOAT      "/sec      0.0654038                                         Yes
29 DEC_rate                       FLOAT      "/sec      -0.028018                                         Yes
30 apparent_magnitude_V           FLOAT      mag        19.16                                             Yes

31 heliocentric_distance_r        DOUBLE     au         1.05402381                                        Yes
32 obscentric_distance_delta      DOUBLE     au         0.28128128                                        Yes
33 lighttime                      FLOAT      sec        140.3607                                          Yes
34 solar_elongation_elong         FLOAT      deg        90.0359                                           Yes
35 before_after_opposition        STRING     ---        /T                                                Yes

36 phase_angle_alpha              FLOAT      deg        -74.49                                            Yes
37 ObsEclLon                      FLOAT      deg         213.2239                                         Yes
38 ObsEclLat                      FLOAT      deg        -3.795164                                         Yes
39 target_X_@sun                  DOUBLE     au          0.319379555249375                                Yes
40 target_Y_@sun                  DOUBLE     au         -1.00430258867518                                 Yes

41 target_Z_@sun                  DOUBLE     au         -0.0185976638442141                               Yes
42 target_X_@observer             DOUBLE     au         -0.235047951829334                                Yes
43 target_Y_@observer             DOUBLE     au         -0.153329693108554                                Yes
44 target_Z_@observer             DOUBLE     au         -0.0186135080544228                               Yes
45 observer_X_@sun                DOUBLE     au          0.554427507078709                                Yes

46 observer_Y_@sun                DOUBLE     au         -0.850972895566624                                Yes
47 observer_Z_@sun                DOUBLE     au          1.58442102086986E-05                             Yes
48 reference_wavelengths_micron   FLOAT      micron     24.0                                               No
49 colour_correction_factor       FLOAT      ---        0.984                                              No
50 colour_corrected_flux_density  DOUBLE     Jy         0.021                                              No

51 absolute_flux_error            DOUBLE     Jy         0.001                                              No
52 comments_remarks               STRING     ---        An assumed geometric albedo of 0.1 was used to     No
                                                        calculate the colour-correction factor.                
53 LTcorrected_epoch              DOUBLE     days       2454308.045366 / to be calculated from             No
                                                        (observation_mid_time - lighttime/3600./24.)	  
54 documents_references           STRING     ---        Mueller T. G. et al. 2014;AKARIAFC                 No
55 date_last_modification         STRING     ---        2018-08-31 15:13:25                                No
56 alt_target_name                STRING     ---        25143#1998 SF36#2025143                            No
\end{verbatim}

\noindent\rule{\textwidth}{0.5pt}
\normalsize
\end{document}